\documentclass[aps,prl,twocolumn,groupedaddress,nofootinbib,notitlepage,showpacs,floatfix]{revtex4-1}

\usepackage{graphicx,graphics,epsfig,times,bm,bbm,amssymb,amsmath,amsfonts,mathrsfs}
\usepackage{color}
\usepackage[pdfstartview=FitH]{hyperref}
\usepackage{natbib}
\usepackage{comment}

\let\oldsqrt\sqrt
\def\sqrt{\mathpalette\DHLhksqrt}
\def\DHLhksqrt#1#2{%
\setbox0=\hbox{$#1\oldsqrt{#2\,}$}\dimen0=\ht0
\advance\dimen0-0.2\ht0
\setbox2=\hbox{\vrule height\ht0 depth -\dimen0}%
{\box0\lower0.4pt\box2}}

\newtheorem{mytheorem}{Theorem}

\newcommand{\ignore}[1]{}
 
\newcommand{\ma}[1]{\textcolor{black}{#1}}

\newcommand{\vs}[1]{\vec{r}_{{{#1}}}}
\newcommand{\vQb}{\vec{r}_{\bar{{Q}}}}

\newcommand{\ket}[1]{ | #1\rangle}
\newcommand{\bra}[1]{\langle #1 | }
\newcommand{\sech}{\textrm{sech}}
\newcommand{\bea} {\begin{eqnarray}}
\newcommand{\eea} {\end{eqnarray}}
\newcommand{\bes} {\begin{subequations}}
\newcommand{\ees} {\end{subequations}}

\excludecomment{oldtext}

\begin{document}

\title{Zeno effect for quantum computation and control}
\author{Gerardo~A. Paz-Silva$^{(1,4)}$, A. T. Rezakhani$^{(1,4,5)}$, {Jason M. Dominy$^{(1,4)}$}, and D.
A. Lidar$^{(1,2,3,4)}$}
\affiliation{Departments of $^{(1)}$Chemistry, $^{(2)} $Physics, and $^{(3)}$Electrical
Engineering, and $^{(4)}$Center for Quantum Information Science \&
Technology, University of Southern California, Los Angeles, California
90089, USA\\
$^{(5)}$Department of Physics, Sharif University of Technology, Tehran, Iran }

\begin{abstract}
It is well known that the quantum Zeno effect can protect specific quantum states
from decoherence by using projective measurements. Here we combine the theory of weak measurements with stabilizer quantum error
correction and detection codes. We derive rigorous performance bounds which demonstrate that the Zeno effect can be used to protect appropriately encoded arbitrary states 
to arbitrary accuracy,
while at the same time allowing for universal quantum computation or quantum control.
\end{abstract}

\maketitle
\affiliation{$^{(1)}$Departments of Chemistry, $^{(2)} $Physics, and $^{(3)}$Electrical
Engineering, and $^{(4)}$Center for Quantum Information Science \&
Technology, University of Southern California, Los Angeles, California
90089, USA\\
$^{(5)}$Department of Physics, Sharif University of Technology, Tehran, Iran }

Protection of quantum states or subspaces of open systems from decoherence
is essential for robust quantum information processing and quantum control.
{The fact} 
that measurements can slow down decoherence is well known as the quantum
Zeno effect (QZE) \cite{Misra:77,Itano:90} (for a recent review see Ref.~\cite{Facchi:08}). The standard approach to the QZE uses repeated strong,
projective measurements of some observable $V$. In this setting{,} 
it can be
shown that such repeated measurements decouple the system from the
environment or bath, and project it into an eigenstate or eigensubspace of $V$ \cite{Facchi:PRL02,Zanardi:1999:77}.
Projective measurements are, however, an
idealization, and here we are interested in the more realistic setting of
weak, non-selective measurements, implementing a weak-measurement quantum
Zeno effect (WMQZE). The measurements are called weak since all outcomes
result in small changes to the state \cite{PhysRevA.41.11,brun:719}, and
non-selective since the outcomes are not recorded. Such measurements can
include many phenomena not captured by projective measurements, e.g.,
detectors with non-unit efficiency, measurement outcomes that include
additional randomness, and measurements that give incomplete information \ma{[see Eq.~\eqref{eq:weak}]}.
The WMQZE has already been considered in a wide range of applications, e.g.,
Refs.~\cite{PhysRevLett.97.260402, Xiao2006424,PhysRevB.73.085317,gong:9984}.
However, a general systematic study of decoherence suppression via the
WMQZE, allowing for universal quantum control, appears to be lacking. This
work aims at bridging this gap. \ma{More specifically, we}%
 ask whether the
WMQZE can be used to protect arbitrary quantum states while they are being
controlled, e.g., for the purpose of quantum computation. Borrowing quantum
coding ideas, we devise a 
measurement protocol 
which allows us to provide an affirmative answer to
this question, namely: Assume the system-bath interaction is local and
bounded, and that we can encode arbitrary system states into a sufficiently
large stabilizer quantum error correcting code. Then a weak
system-measurement protocol using $M$ stabilizer measurements of strength $\epsilon $ lasting a total time $\tau $ suppresses the system-bath
interaction arbitrarily well 
in the limit of large 
$M$,
and commutes with quantum control or
computation being performed on the system. We dedicate the rest of this work
to explaining, sharpening, and proving this claim.

\textit{Weak measurements}.---A generalized, positive
operator-valued measure (POVM)\ comprises a set of \textquotedblleft
measurement operators\textquotedblright\ $\{M_{j}\}$ satisfying the sum rule 
$\sum_{j}M_{j}^{\dag }M_{j}=\openone$, which map a state $\varrho $ to $\varrho _{j}=M_{j}\varrho M_{j}^{\dag }/p_{j}$ with probability $p_{j}=\mathrm{Tr}[M_{j} \varrho M_{j}^{\dag }]$ {for measurement}
outcome $j$ \cite{Nielsen:book}. In general, one can write the 
\emph{weak measurement superoperator} corresponding to a two-outcome
measurement of an observable (Hermitian operator) $V$ with strength $\epsilon $ on a state $\varrho $ as \cite{Oreshkov:05a}: 
${\mathcal{P}}_{\epsilon}(\varrho ) = \sum_{r=\pm }P_{V}(r\epsilon )\varrho
P_{V}(r\epsilon )$, where $P_{V}(\epsilon ) =\sum_{s=\pm }\alpha
_{s}(\epsilon )P_{sV}$, with $P_{\pm V} \equiv \frac{1}{2}(\openone\pm V)$
standard projection operators when $V^2=\openone$, and 
$\alpha _{\pm }(\epsilon ) \equiv \sqrt{{(1\pm \tanh (\epsilon )})/{2}}$.
Since $P_{V}^{2}(\epsilon )+P_{V}^{2}(-\epsilon )=\openone$, the operators $\{P_{V}(\epsilon ),P_{V}(-\epsilon )\}$ satisfy the sum rule, and hence are
measurement operators for a given $\epsilon $. They are parametrized by the
strength $\epsilon $ so that they can be considered 
\emph{weak measurement operators}. Since $\lim_{\epsilon \rightarrow \pm
\infty }P_{V}(\epsilon )=P_{\pm V}$, the ideal or strong measurement limit
is recovered when the measurement strength $|\epsilon| \rightarrow \infty $,
i.e., ${\mathcal{P}}_{\infty }(\varrho)= 
\sum_{s=\pm }P_{sV}\varrho P_{sV}$. The no-measurement scenario
is
the case $\epsilon \rightarrow 0$, i.e., ${\mathcal{P}}_{0}(\varrho)=\varrho $. 
The weak measurement of an operator $V$ with strength $\epsilon$ can be rewritten as%
\ma{
\bea
\mathcal{P}_{\epsilon} (\varrho) = (1- \zeta)  \mathcal{P}_{\infty} (\varrho) + \zeta \varrho, \quad \zeta\equiv \sech(\epsilon),
\label{eq:weak}
\eea
}%
and thus a weak measurement can be interpreted as a noisy measurement in which, with probability 
\ma{$\zeta$}, 
the measurement is not executed \cite{supp}. A strong measurement is the idealized case, when 
\ma{$\zeta=0$}.
Weak measurements are universal
in the sense that they can be used to build
up arbitrary measurements without the use of ancillas \cite{Oreshkov:05a}. 

\textit{Open system evolution with measurements}.---Consider a system and
bath with respective Hilbert spaces $\mathcal{H}_{S}$ and $\mathcal{H}_{B}$.
The joint evolution is governed by the Hamiltonian $H = H_{0}+H_{SB}$, where 
$H_{0} \equiv H_{S} \otimes \openone_{B}+\openone_{S}\otimes H_{B}$, acting
on the joint Hilbert space $\mathcal{H}_{SB} \equiv \mathcal{H}_{S}\otimes 
\mathcal{H}_{B}$. We assume that $\Vert H_{\mu }\Vert {\equiv} J_{\mu }/2<\infty$
($\mu\in\{0,S,B,SB\}$) \cite{norms}.
We denote $J_1 \equiv J_{SB}$.
Thus $\Vert H\Vert \leq \left( J_{0}+J_1\right) /2\equiv J/2<\infty $. 

We wish to protect an arbitrary and unknown system state $\varrho _{S}$ 
against decoherence for some time $\tau $ using only weak measurements. We
model all such measurements as instantaneous and perform $M$ equally-spaced
measurements in the total time $\tau $. We define superoperator generators ${\mathcal{L}}_{\mu }(\cdot )\equiv -i[H_{\mu },\cdot ]$ and ${\mathcal{L}}(\cdot )\equiv -i[H,\cdot ]$. The free evolution superoperator ${\mathcal{U}}(\tau )\equiv e^{{\mathcal{L}}\tau }$ describes the evolution after each
measurement. Hence, the joint state and system-only state, after time $\tau$, are given by 
\begin{equation}
\hskip-2mm
\varrho _{SB}(\tau )=\left( {\mathcal{P}}_{\epsilon }{\mathcal{U}}(\frac{\tau}{M})\right) ^{M}\!\!\varrho _{SB}(0), \, \varrho _{S}(\tau )=\mathrm{Tr}_{B}\varrho _{SB}(\tau ) .
\label{wQZ}
\end{equation}
From now on we shall assume for simplicity
that the initial system state is pure: $\varrho _{S}(0)=|\psi _{S}(0)\rangle \langle \psi_{S}(0)|$ and that 
the joint initial state is factorized, i.e., $\varrho _{SB}=\varrho _{S}\otimes \varrho _{B}$. 
For notational simplicity we denoted 
$\varrho _{\mu}(0)\equiv \varrho _{\mu}$.
Note that in Eq.~(\ref{wQZ}) ${\mathcal{P}}_{\epsilon }$ acts
non-trivially only on system operators.

\textit{Figure of merit}.---To determine the success of our protection
protocol we compare the \textquotedblleft real\textquotedblright\ system
state with protection and in the presence of $H_{SB}$ [Eq.~(\ref{wQZ})] to
the uncoupled ($H_{SB}=0$), unprotected \textquotedblleft
ideal\textquotedblright\ system state, namely to $\varrho _{S}^{0}(\tau )=\mathrm{Tr}_{B}
\varrho _{SB}^{0}(\tau )$, with $\varrho _{SB}^{0}(\tau )={\mathcal{U}}_{0}(\tau )\varrho _{SB}$, where ${\mathcal{U}}_{0}(\tau )\equiv
e^{{\mathcal{L}}_{0}\tau }$ and$~{\mathcal{L}}_{0}={\mathcal{L}}_{S}+{\mathcal{L}}_{B}$ ($[{\mathcal{L}}_{S},{\mathcal{L}}_{B}]=0$) are the
\textquotedblleft ideal\textquotedblright\ unitary superoperator and its
generator, respectively.

A suitable figure of merit is then the trace-norm distance 
\cite{norms,Nielsen:book} $D[\varrho_1,\varrho_2] \equiv \frac{1}{2}\|
\varrho_1-\varrho_2\| _{1}$ between the real and ideal states.
We shall show that we can make $D[\varrho_S(\tau),\varrho_S^0(\tau)]$  arbitrarily small for a given $H$ by a suitable choice of weak measurements.

\textit{Weak measurements over a stabilizer code}.---Previous 
WMQZE work
applied only to particular states
\cite{PhysRevLett.97.260402, Xiao2006424,PhysRevB.73.085317,gong:9984}. 
To achieve our goal of protecting
an arbitrary, unknown $k$-qubit state, 
we encode
the state into an $[[n,k,d]]$ stabilizer quantum error correcting code
(QECC) \cite{Gottesman:96,Nielsen:book}, with stabilizer group $\mathbf{S}=\{S_{i}\}_{i=0}^{Q}$, and where $S_{0}\equiv \openone$.
We assume that
the code distance $d\geq 2$, i.e., the code is at least error-detecting,
with generators $\mathbf{\bar{S}=\{}\bar{S}_{i}\}_{i=1}^{\bar{Q}}\subset 
\mathbf{S}$, 
where $\bar{Q}=n-k$.
Note that every stabilizer element can be written as $S_{i}=\prod_{\nu =1}^{\bar{Q}}\bar{S} _{\nu }^{r_{i\nu }}$, where ${r_{i\nu }}\in \{0,1\}$,
i.e., the stabilizer elements are given by all possible products of the
generators, whence $Q+1=2^{\bar{Q}}$.
The encoded initial state $|\psi _{S}(0)\rangle $ is a simultaneous $+1$ eigenstate of all the elements
of $\mathbf{S}$. We can associate a pair of projectors (measurement
operators) $P_{\pm S_{i}}\equiv \frac{1}{2}(\openone\pm S_{i})$ to each
stabilizer group element, and accordingly a pair of 
weak measurement operators $\{P_{S_{i}}(\epsilon),P_{S_{i}}(-\epsilon )\}$ to each $S_{i}$, i.e., $P_{S_{i}}(\epsilon)=\sum_{s=\pm }\alpha _{s}(\epsilon)P_{sS_{i}}$. 
\ma{In quantum error correction (QEC) one performs a strong measurement of the generators in order to extract an error syndrome \cite{Gottesman:96}. It has been recognized that these strong syndrome measurements implement a QZE \cite{PhysRevA.54.R1745,PhysRevA.72.012306}. 
When we measure $\mathbf{\bar{S}}$ we need to form
products of the weak measurement operators of all the generators, accounting
for all possible sign combinations.%
Let $P_{\mathbf{\bar{S}}}^{(b)}(\epsilon )\equiv \prod_{i=1}^{\bar{Q}}P_{\bar{S}_{i}}\left( (-1)^{b_{i}}\epsilon \right) $} 
denote
such a product for a given choice of signs uniquely determined by the
integer $b=\sum_{i=0}^{\ma{\bar{Q}}}b_{i}2^{i}$, with $b_{i}\in \{0,1\}$. 
\ma{Letting
\begin{equation}
\overline{\mathcal{P}}_{\epsilon}(\varrho) =\sum_{b=0}^{2^{\bar{Q}}-1}P_{\mathbf{\bar{S}}}^{(b)}(\epsilon
)\varrho P_{\mathbf{{\bar{S}}}}^{(b)}(\epsilon )
\label{compmeas}
\end{equation}
we can now define a  \emph{weak stabilizer generator measurement protocol} as $\left(\overline{{\mathcal{P}}}_{\epsilon }{\mathcal{U}}(\frac{\tau}{M})\right) ^{M}$.}

\ma{We stress the two important differences between this protocol and the analogous stabilizer measurement step in QEC: first, we do not need to observe or use the syndrome; second, we allow for weak measurements. In this sense our assumptions are weaker than those of QEC, and hence the ability to perform QEC implies the ability to perform our protocol.}

\ma{Moreover, for the same reason that the many-body character of stabilizer measurements is not a significant drawback in QEC theory, it is not a problem for our protocol either. The reason is that such measurements can be implemented (even fault-tolerantly) using at most two-local operations. See \cite{supp} for the explicit two-local construction for  the weak measurement case. An alternative is to consider a protocol based on measuring the gauge operators of the Bacon-Shor code \cite{Bacon:2006:012340}, which are all two-local, and can be shown to implement a WMQZE as well \cite{Paz-tbp}.}

\ma{{We shall also consider a \emph{weak stabilizer group measurement protocol}:} {$\left({{\mathcal{P}}}_{\epsilon }{\mathcal{U}}(\frac{\tau}{M})\right) ^{M}$,} {where 
${\mathcal{P}}_{\epsilon } (\varrho)=$} {$\sum_{b=0}^{2^{Q}-1}P_{\mathbf{S}}^{(b)}(\epsilon )\varrho P_{\mathbf{S}}^{(b)}(\epsilon )$,} {with $P_{\mathbf{S}}^{(b)}(\epsilon
)\equiv \prod_{i=0}^{Q}P_{S_{i}}\left( (-1)^{b_{i}}\epsilon \right) $  and $b=\sum_{i=0}^{{{Q}}}b_{i}2^{i}$.}
{As we shall see, the generators and group protocols exhibit substantial tradeoffs,} so we shall consider both
in our general development below.}

Note that if $\varrho _{S}$ is stabilized by
\ma{$\bar{\mathbf{S}}$ (or ${\mathbf{S}}$)} then
the weak measurement protocol perfectly preserves 
an arbitrary
encoded state in the absence of system-bath coupling.
Another
important fact we shall need later is that given some $[[n,k,d]]$
stabilizer QECC, if a Pauli group operator $P$ anticommutes with at least
one of the stabilizer generators, then it anticommutes with half of all the
elements of the corresponding stabilizer group $\mathbf{S}$ \ma{\cite{supp}}.%

\textit{Distance bound}.---
Following standard conventions, we call a Pauli operator $k$-local if it
contains a tensor product of $k$ non-identity Pauli operators. We call a
system Hamiltonian $k$-local if it is a sum of $k$-local Pauli operators,
and a system-bath Hamiltonian $k$-local if it is a sum of $k$-local Pauli
operators acting on the system, tensored with arbitrary bath operators. 

{Let $\vec{r}_s = \{r_i\}_{i=1}^{s}$, where $r_i\in\{0,1\}$ $\forall i$ and $s\in\{1,\dots \bar{Q}\}$. Let $\{ \Omega_i\}_{i=1}^{\bar{Q}}$, where $\Omega_i^2 = \openone$ $\forall i$, denote a commuting set of operators acting on the system only. Consider the recursive definition 
$H_{\vs{s}} = \frac{1}{2}\left(H_{\vs{s-1}}+(-1)^{r_s}\Omega_s H_{\vs{s-1}} 
\Omega_s\right)$, where $H_{\vs{0}} \equiv H$.
This construction 
allows for the decomposition of any Hamiltonian as 
$H= \sum_{\vQb}H_{\vQb}$, with the property $\{H_{\vQb} , 
\Omega_i\} = 0$  if $r_i=1$, or  $[H_{\vQb} , \Omega_i] = 0$  if $r_i=0$. Note that $H_0 = H_{\vec{0}_{\bar{Q}}}$
and 
\ma{$H_{SB} = \sum_{\vQb} H_{\vQb}-H_0$}. 
It follows from the 
triangle inequality, norm submultiplicativity, and the 
recursive definition of $H_{\vs{i}}$ that $\Vert H_{\vQb} 
\Vert \leq J_{1}/2$. 
These bounds can be further specialized or tightened for specific forms 
of the Hamiltonian.} 

We are now ready to state our main result:
\begin{mytheorem}
\label{Th:1} 
Assume an arbitrary pure state $\varrho_S = |\psi _{S}\rangle \langle \psi_S|$ 
is encoded into an $[[n,k,d]]$ stabilizer QECC. {Assume that $H_{SB}=
\sum_{K=1}^{d-1}H_{SB}^{(K)}$ and that $H_{S}$
commutes with the code's stabilizer, so that $H_{S}=\sum_{l\geq 1}H_{S}^{(ld)}$},
where 
$H_{SB}^{(K)}$ ($H_{S}^{(K)}$) denotes a $K$-local system-bath (system-only)
Hamiltonian, and all Hamiltonians, including $H_{B}$, are bounded in the 
sup-operator norm. {Finally,} let $Q=2^{n-k}-1$ and $q=(Q+1)/2$, {and assume $J_{0}>J_{1}$}.
Then 
the stabilizer group measurement protocol 
$\left({\mathcal{P}}_{\epsilon}{\mathcal{U}}(\tau /M)\right) ^{M}$ protects $\varrho_S$ 
up to a deviation 
that 
converges 
to $0$ in the large $M$ limit:
\begin{subequations}
\label{eq:D-bound}
\begin{align} 
&D[\varrho _{S}(\tau ),\varrho _{S}^{0}(\tau )] \leq A_{+}\gamma_{+}^{M-1} + A_{-}\gamma_{-}^{M-1} - e^{{J_0} \tau} {\equiv B}
\label{full-bound}\\
& = \frac{Q}{2}e^{\tau J_{0}}\!\!\left[\frac{\tau^{2}J_{1}^{2}}{2}\! + \!\big(\tau J_{0} \!+ \!\tau^{2} J_{0}^{2}\big)\frac{\zeta^{q}}{1\!-\!\zeta^{q}}\right]\!\!\frac{1}{M}\! +\! O\left(\!\frac{1}{M^{2}}\!\right)
\label{eqn:asymptoticbound}
\end{align}
\end{subequations}
where 
\begin{subequations}
\label{eq:details}
\begin{eqnarray}
	\beta & \equiv & {e^{\frac{\tau J_{0}}{M}}\left[\frac{Qe^{-\frac{\tau J_{1}}{M}} + e^{\frac{\tau J_{1} Q }{M}}}{Q+1}\right] - 1} 
	\label{eqn:beta}	\\
	\gamma_{\pm} & \equiv & \frac{1}{2}\big(1+\beta + (1+Q\beta)\zeta^{q}\big) \nonumber\\
	& &  \pm \frac{1}{2}\sqrt{\big(1+\beta-(1+Q\beta)\zeta^{q}\big)^{2} + 4 Q \beta^{2}\zeta^{q}}
	\label{eqn:rootspm}\\
	A_{\pm} & \equiv & \frac{Q\beta\zeta^{q}(\gamma_{\pm}+\beta) + (1+\beta)\big[(1+\beta)-\gamma_{\mp}\big]}{\gamma_{\pm}-\gamma_{\mp}}
	\label{eqn:coeffspm}.
\end{eqnarray}
\end{subequations}
For a generator measurement protocol $\left({\overline{\mathcal{P}}}_{\epsilon }
{\mathcal{U}}(\tau /M)\right)^{M}$, replace 
$q$ by 1 {in Eqs.~(\ref{eqn:asymptoticbound}), \eqref{eqn:rootspm}, and (\ref{eqn:coeffspm}).  In the strong measurement limit ($\epsilon\to\infty$), both protocols yield the distance bound 
\begin{equation}
D[\varrho _{S}(\tau ),\varrho _{S}^{0}(\tau )] \leq e^{J_0 \tau}\!\left[ \left( \frac{Q e^{-\frac{J_1\tau}{ M}}\!+e^{\frac{J_1 \tau Q }{M}}}{Q+1}\right)^M \!\!\!\! -1 \right].
\label{eq:strong}
\end{equation}}
\end{mytheorem}

To motivate the locality aspects of Theorem~\ref{Th:1} recall that by
construction of a stabilizer code any Pauli operator with locality $\leq d-1$
anticommutes with at least one stabilizer generator, a condition satisfied
by all $H_{SB}^{(K)}$ in Theorem~\ref{Th:1}. Moreover, logical operators of
the code (elements of the normalizer, which commute with the stabilizer)
must have locality that is an integer multiple of the code distance $d$, a
condition satisfied by every $H_{S}^{(ld)}$, which
by assumption
can be used to implement logical operations on the code while stabilizer 
measurements are
taking place. To keep the locality of $H_{S}$ low thus requires a low
distance code. We present an example of a $d=2$ code below.

\textit{Proof sketch of Theorem \ref{Th:1}}.---We first consider the case of
weak measurements of the entire stabilizer group, ${\mathcal{P}}_{\epsilon} $. A typical $K$-local term we need to calculate is then of the form ${\mathcal{P}}_{\epsilon }(H_{SB}^{(K)}\varrho _{SB})=\sum_{b=0}^{2^{Q}-1}P_{\mathbf{S}}^{(b)}(\epsilon )H_{SB}^{(K)}\varrho _{SB}P_{\mathbf{S}}^{(b)}(\epsilon )$. Now we use the previously established fact that if $E$
(modulo logical operators and stabilizer operations) is a correctable error,
then $\{S_{i},E\}=0$ for exactly half of the stabilizer elements. Hence the
same number of stabilizer elements, $q=(Q+1)/2$, anticommute with $H_{SB}^{(K)}$. From here a straightforward calculation 
reveals that $({\mathcal{P}}_{\epsilon })^{j}{\mathcal{L}}_{SB}(\varrho _{SB})=\zeta^{jq}{\mathcal{L}}_{SB}(\varrho _{SB})$, a key result since it shows how the measurements suppress the ``erred" component of the state, ${\mathcal{L}}_{SB}(\varrho _{SB})$. On the other hand, since we
assume that $[H_{S},S_{i}]=0$ for all stabilizer elements we have ${\mathcal{P}}_{\epsilon }\big({\mathcal{L}}_{0}(\varrho _{SB})\big)={\mathcal{L}}_{0}(\varrho _{SB})$ 
and hence $({\mathcal{P}}_{\epsilon })^{j}\big({\mathcal{L}}_{0}(\varrho _{SB})\big)={\mathcal{L}}_{0}(\varrho _{SB})$, meaning that measurements do not interfere
with the ``ideal'' evolution. 

Taylor expanding $\mathcal{U}(\tau/M) = \exp[(\tau/M)\mathcal{L}]$ in Eq.~(\ref{wQZ}) and the ÒidealÓ unitary superoperator
$\mathcal{U}_{0}(\tau)$, and expanding $\mathcal{L}$ as a sum of $K$-local terms yields an expression for $\varrho_{SB}(\tau) - \varrho_{SB}^{0}(\tau)$ as a sum of products of projectors $\mathcal{P}_{\epsilon}$ and Hamiltonian commutators $\mathcal{L}_{\vQb}(\cdot ) \equiv -i[H_{\vQb},.]$ acting on $\varrho_{SB}$.  By the above arguments, the projectors in each of these terms may be replaced by $\zeta^{jq}$, where $j= 0$ if 
all commutators in the term are $\mathcal{L}_{0} \equiv \mathcal{L}_{\vec{0}_{\bar{Q}}}$. 
Invoking the triangle inequality, submultiplicativity, and the fact that 
\ma{$\|\varrho_{SB}\|_1=1$},
allows the trace-norm of this sum to be bounded by a linear combination of norms of the ${\mathcal{L}_{\vQb}}$ operators, which may all then be replaced by the upper bounds $J_{0}\geq{\|\mathcal{L}_{0}\|}$
and $J_{1}\geq{\|\mathcal{L}_{\vQb}\|}$
or all ${\vQb}\neq \vec{0}_{\bar{Q}}$.  The resulting \ma{hypergeometric} sum may be shown to equal the expression 
\ma{$B$} given in Eq.~(\ref{full-bound}).

When we perform generator measurements ${\overline{\mathcal{P}}}_{\epsilon}$,
each error anticommutes with at least one generator. To 
derive a simple
but general result we only consider the worst case 
scenario of each error anticommuting with just one generator.
An almost identical calculation to
the one for 
the full stabilizer group protocol reveals an upper bound for $D[\varrho_S(\tau),\varrho_S^0(\tau)]$ given by replacing $q$ by $1$ in Eqs.~\ma{\eqref{eq:D-bound} and \eqref{eq:details}},
since $q$ 
counts the number of anticommuting stabilizer or generator elements. This completes the proof sketch. Complete proof details will be provided in Ref.~\cite{future}.

\begin{figure*}
\centering
\includegraphics[width=17cm]{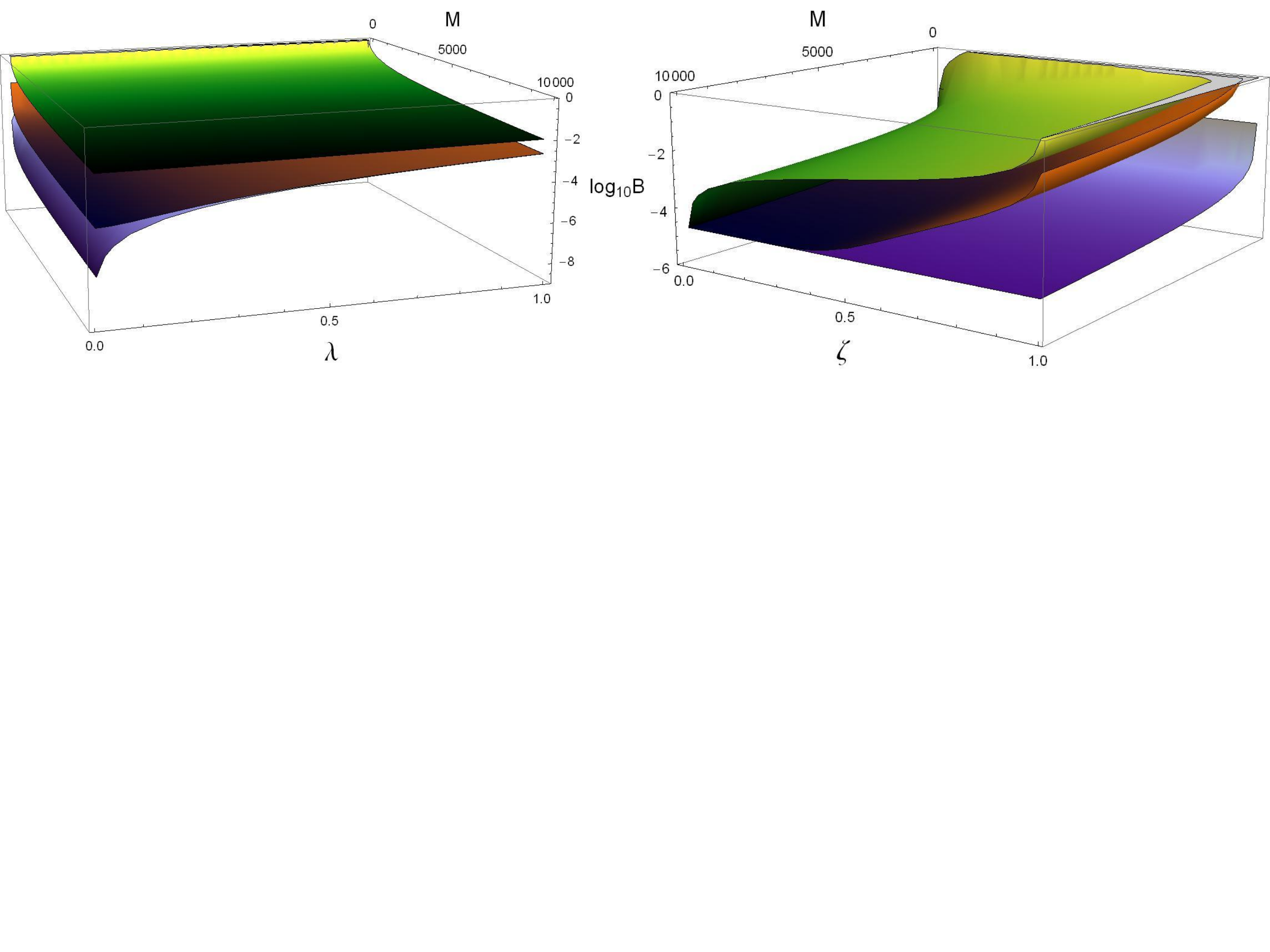}
\vspace{-7cm}
\caption{{(Color online) Left: the upper bound $B$ [Eq.~\eqref{full-bound}] as a function of the number of measurements $M$ and $\lambda \equiv J_1/J_0$, with $J_0 \tau=1$, $\bar{Q}=4$ and {$\zeta=0.5$}. Right: the same bound as a function of $M$ and $\zeta$, with $J_0 \tau=1$, $\bar{Q}=4$, and $\lambda=0.1$. In both plots the upper, middle, and lower surfaces are, respectively, the bounds for the generators-only, full stabilizer-group, and strong measurements protocols, the latter being the $\zeta \rightarrow 0$ limit of $B$, given in Eq.~\eqref{eq:strong}. The full stabilizer-group bound is tighter than the generators-only bound for all values of the parameters, and is closer to the bound for the strong measurement limit.}}
\label{figtest-fig}
\end{figure*}

We note that the generators-only bound is not as tight as the one for the full-group protocol, due to the worst case assumption of $q=1$ used to upper-bound terms with larger exponents which appear in the Taylor expansion discussed above. I.e., the bound on the generators-only protocol contains a sum over terms of the form $\mathcal{P}_\epsilon \mathcal{L}_{\vQb} (\varrho)= \zeta^{q} \mathcal{L}_{\vQb} (\varrho)$, where $q\in\{1,\dots,\bar Q\}$, all of which we have replaced for simplicity by $q=1$. Our upper bounds are illustrated in Fig.~\ref{figtest-fig}. \ma{Clearly, the generators-only bound is not as close to the strong measurement limit as the full-group protocol bound. However, the former protocol requires an exponentially smaller number of measurements.  If the measurement is performed, e.g., by attaching an ancilla for each measured Pauli observable (as in a typical fault-tolerant QEC implementation \cite{Nielsen:book}), then this translates into an exponential saving in the number of such ancillas. Thus the two protocols exhibit a performance-resource tradeoff. Next, we discuss an example}.

\textit{Suppression of 1-local errors}.---To illustrate 
our general construction we consider suppression of decoherence due to
a Hamiltonian containing $1$-local errors on $n$ qubits: $
H_{SB}=\sum_{i=1}^{n}\sum_{\alpha \in \{x,y,z\}}\sigma _{i}^{\alpha }\otimes
B_{i}^{\alpha }\equiv H_{x}+H_{y}+H_{z}$, where $J_{a}\equiv \|H_a\|<\infty$. 
This model captures the dominant errors in any implementation of quantum
control or quantum computing using qubits, since any terms with higher
locality must result from 3-body interactions and above. Theorem \ref{Th:1}
guarantees first order suppression of this $H_{SB}$ provided we perform weak
measurements over a stabilizer group of distance $d\geq 2$. We can, e.g.,
choose an error detection code $\mathcal{C}=[[n,n-2,2]]$, where $n $ is
even, defined by the stabilizer generators $\mathbf{\bar{S}}=\{\bar{S}
_{1}=X^{\otimes n},\bar{S}_{2}=Z^{\otimes n}\}$, i.e., $\bar{Q}=2$. The
codewords are $\{|\psi _{x}\rangle =\left( |x\rangle +|\bar{x}\rangle
\right) /\sqrt{2}\}$, where $x$ is an even-weight binary string of length $n $ and 
$x+\bar{x}=0 \pmod 2$. 
This distance $d=2 $ code 
is attractive since the normalizer elements are all 
$2$-local, which means that $H_{S}$ is also $2$-local if it is constructed
over these normalizer elements. Encoded single-qubit operations for $\mathcal{C}$
 are $\tilde{X}_{j}=\sigma _{1}^{x}\sigma _{j+1}^{x}$ and 
 $\tilde{Z}_{j}=\sigma _{j+1}^{z}\sigma _{n}^{z}$ , where $j=1, \ldots, n-2$.
Encoded two-qubit interactions are $\tilde{X}_{i}\tilde{X} _{j}=\sigma
_{i+1}^{x}\sigma _{j+1}^{x}$ and $\tilde{Z}_{i}\tilde{Z}_{j}=\sigma
_{i+1}^{z}\sigma _{j+1}^{z}$. This is sufficient for universal quantum
computation in both the circuit \cite{Nielsen:book} and adiabatic models 
\cite{Biamonte:07}. Thus our encoded WMQZE strategy applies in both settings (we note that our results hold even when the Hamiltonian $H$ is time-dependent \cite{future}).
If we weakly measure the entire stabilizer group $\mathbf{S}=\{\openone
,X^{\otimes n},Y^{\otimes n},Z^{\otimes n}\}$, Theorem \ref{Th:1} implies 
that a state encoded into $\mathcal{C}$, supporting $n-2$ logical qubits, {is protected by the encoded WMQZE according to Eq.~\eqref{eqn:asymptoticbound} with $Q=3$ and $q=2$.}
\ignore{
\begin{align}
	&D[\varrho _{S}(\tau ),\varrho _{S}^{0}(\tau )]\leq \nonumber\\
	& \frac{3}{2}e^{\tau J}\!\left[\frac{\tau^{2}J^{2}}{2}\! + \!\big(\tau J \!+ \!\tau^{2} J^{2}\big)\frac{\zeta^{2}}{1-\zeta^{2}}\right]\!\frac{1}{M}\! +\! O\left(\!\frac{1}{M^{2}}\!\right).
\end{align}
}
If we measure only the generators,
Theorem \ref{Th:1} gives the same bound with $\zeta ^{2}$ replaced by $\zeta$.

\textit{Conclusions}.---The \textquotedblleft traditional\textquotedblright\
QZE uses strong, projective measurements, and is only able to protect an
eigenstate of the operator being measured. In this work we have presented
a 
general study of decoherence suppression via the WMQZE for
arbitrary quantum states, allowing for universal quantum 
computation and
control. By using the WMQZE to protect codewords of a stabilizer QECC, we have 
explicitly demonstrated that one can achieve decoherence suppression to arbitrary 
accuracy by increasing both the measurement strength and frequency, while at the
same time applying logical operators (normalizer elements) as Hamiltonians,
which suffices for universality. This establishes the WMQZE as a general
alternative to other open-loop quantum control methods, appropriate where
measurements, rather than unitary control, is advantageous. A natural
example is measurement-based quantum computation \cite{Raussendorf:01}. We
defined two protocols, one based on measurement of the full stabilizer
group, another on measurement of the generators only, and studied the
tradeoff between the two. The former requires exponentially more commuting
measurements. However, \ma{our upper bound on its suppression of} 
\ma{the effect of the finiteness of the measurement strength is} exponentially \ma{tighter}.

It would be interesting to consider whether---similarly to recent developments in dynamical decoupling theory using concatenated sequences \cite{KhodjastehLidar:04} or pulse interval
optimization \cite{Uhrig:07,WFL:09,NUDD}---WMQZE decoherence suppression can be optimized by exploiting, e.g., recursive design or non-uniform measurement intervals. Another interesting possibility is to analyze the joint effect of feedback-based quantum error correction \cite{Gottesman:96} and the encoded WMQZE. \ma{Finally, it would be interesting to improve the WMQZE protocol using techniques from fault tolerance theory \cite{Aliferis:05}.}

\textit{Acknowledgment}.---DAL acknowledges support from the U.S. Department
of Defense and the NSF under Grants No. CHM-1037992 and CHM-924318. 


\begin{thebibliography}{29}
\bibitem [{nor()}]{norms}%
  \BibitemOpen
  \href@noop {} {}\bibinfo {note} {For any linear operator $A$ with singular
  values $s_{i}(A)$ (eigenvalues of $\sqrt{A^{\dagger }A}$), $\Vert A\Vert
  \equiv \sup_{i}s_{i}(A)$ and $\Vert A\Vert _{1}\equiv
  \mathrm{Tr}\sqrt{A^{\dagger }A}=\sum_{i}s_{i}(A)$ are the standard
  sup-operator norm and trace norm, respectively.}\BibitemShut {Stop}%
\end{thebibliography}

%

\widetext

\begin{center}
\Large{Supplementary Material}
\end{center}

\section{Weak measurements}

{
Recall that $\alpha _{\pm }(\epsilon ) \equiv \sqrt{{(1\pm \tanh (\epsilon )})/{2}}$. 
Note the identities 
\bes
\begin{eqnarray}
\alpha _{\pm }^{2}(\epsilon )+\alpha _{\pm }^{2}(-\epsilon ) &=&1,
\label{id1} \\
\alpha _{+}(\epsilon )\alpha _{-}(\epsilon )+\alpha _{+}(-\epsilon )\alpha
_{-}(-\epsilon ) &=&\mathrm{sech}(\epsilon )\equiv \zeta .  \label{id2}
\end{eqnarray}
\ees
}
Now consider the expression for a measurement of strength $\epsilon$ of an operator $V$.
{Using Eqs.~\eqref{id1}, \eqref{id2}, and $P_{\pm V} \equiv \frac{1}{2}(\openone\pm V)$, we have:}
\bes
\label{eq:weak-expanded}
\begin{eqnarray} 
\mathcal{P}_{\epsilon} (\varrho) &=& \sum_{r=\pm} \sum_{s,s'=\pm} \alpha_s (r \epsilon) P_{sV} \varrho \alpha_{s'} (r \epsilon) P_{s'V}	 \\
&=& \sum_{r=\pm}  \left(\alpha^2_+ (r \epsilon) P_{+V} \varrho P_{+V} + \alpha_+ (r \epsilon) \alpha_{-} (r \epsilon)  P_{+V} \varrho P_{-V} + \alpha_- (r \epsilon) \alpha_{+} (r \epsilon)  P_{-V} \varrho P_{+V} +\alpha^2_- (r \epsilon)   P_{-V} \varrho P_{-V}\right)\\
&=& P_{+V} \varrho P_{+V} + P_{-V} \varrho P_{-V} + \zeta \left( P_{+V} \varrho P_{-V} + P_{-V} \varrho P_{+V}\right)\\
&=& (1-\zeta) \left(P_{+V} \varrho P_{+V} + P_{-V} \varrho P_{-V}\right) + \zeta \varrho\\
&=& (1-\zeta) \mathcal{P}_{\infty}{(\varrho)} + \zeta \mathcal{P}_{0}{(\varrho)} ,
\end{eqnarray}
\ees
{which is a convex combination of the no-measurement map $\mathcal{P}_{0}$ and the strong measurement map $\mathcal{P}_{\infty}$. Thus a weak measurement is a measurement which allows for the strong measurement not having taken place with probability $\zeta$. This could be due to detectors with non-unit efficiency, measurement outcomes that include
additional randomness, and measurements that give incomplete information. Strong measurements, i.e., $\zeta =0$, are therefore an idealization.}

\section{One anticommutation implies more}
In the paper we stated that if a Pauli group operator $P$ anticommutes with at least one of the stabilizer generators, then it anticommutes with half of all the elements of the corresponding stabilizer group. To prove this
claim consider the generator subset $\bar{C}(P)\equiv \{\bar{S}_{j},j =1,
\ldots, \kappa\leq \bar{Q}:\{\bar{S}_{j},P\}=0\}$, for a fixed Pauli
operator $P$. Then any stabilizer element $S_{i}$ resulting from a product
of an odd number of the members of $\bar{C}(P)$ anticommutes with $P$; let
us group all those elements in the set $C(P)$. We now show that for every
element of $C(P)$ there is an element not in $C(P)$ which belongs to $\mathbf{S}$. If we multiply each element of $C(P)$ by one fixed member of $\bar{C}(P)$, say $\bar{S}_{1}$, we obtain a set of elements which do not
belong to $C(P)$, the set $C^{\prime }(P)$ which has the same number of
elements. Similarly, multiplying each element $S_{i}\notin C(P)$ by $\bar{ S}_{1}$ maps each element to $C(P)$. Since $\mathbf{S}$ is a group it follows
that $\mathbf{S}=C(P)\cup C^{\prime }(P)$, where $|C(P)|=|C^{\prime }(P)|$, 
as required. 

\section{Two-body implementation of many-body weak measurements}

{In order to implement the many-body weak-measurements using only physically reasonable one- or two-body operations, we can use a standard construction from from fault-tolerance theory \cite{Gottesman:1998:127}. As an added benefit, this construction is fault-tolerant, i.e., errors are not propagated in a harmful way.}

{Consider the measurement of a $k$-local many-body Pauli operator  $\hat{V}=V_1 \otimes \cdots \otimes V_k$, where $V_i \in\{\openone, X, Z, Y\}$. The protocol $\left(\mathcal{P}_{\epsilon,\hat{V}}{\mathcal{U}}(\frac{\tau}{M})\right) ^{M}(\varrho)$ means that we apply $M$ non-selective measurements $\mathcal{P}_{\epsilon,\hat{V}}$ separated by time-intervals $\frac{\tau}{M}$. Our goal is to show that each of these measurements can be simulated using only single-qubit measurements and $2$-qubit gates.}

{When we Taylor-expand $\mathcal{U}(\frac{\tau}{M})$, the terms that arise from the products of Hamiltonians (the generators of $\mathcal{U}$) are all of the form  $ \sum_{\alpha, \beta \in\{0,1\}} H_{\alpha} \varrho H_\beta$, where $H_0$ and $H_1$ group those sums of products of Hamiltonian terms that commute or anticommute with $\hat{V}$, respectively, and where $\hat{V}$ stabilizes $\varrho$. 
We shall show that the action of $\mathcal{P}_{\epsilon, \hat{V}}$ on each term $ \sum_{\alpha, \beta \in\{0,1\}} H_{\alpha} \varrho H_\beta$ can be simulated using only single-qubit measurements in the $Z$-basis and $2$-qubit controlled-NOT ($CX$) and controlled-phase ($CZ$) gates. By linearity, this will imply the result for the entire protocol $\left({{\mathcal{P}}}_{\epsilon,\hat{V}}{\mathcal{U}}(\frac{\tau}{M})\right) ^{M}(\varrho)$.}

{
\subsection{The many-body weak measurement}
Let us first discuss the outcome of a single instance of the many-body weak measurement. Recall that ${\mathcal{P}}_{\infty,\hat{V}}(\varrho)= 
\sum_{s=\pm }P_{s\hat{V}}\varrho P_{s\hat{V}}$, with $P_{\pm \hat{V}} \equiv \frac{1}{2}(\openone\pm \hat{V})$. Clearly, $\mathcal{P}_{\infty, \hat{V}} \left( H_{\alpha} \varrho H_\beta  \right) = 0$ if either $H_\alpha$ or $H_\beta$ (but not both) anticommutes with $\hat{V}$, and $\mathcal{P}_{\infty, \hat{V}}\left( H_{\alpha} \varrho H_\beta  \right)= H_{\alpha} \varrho H_\beta$ if both $H_\alpha, H_\beta$ or neither $H_\alpha,H_\beta$ anticommute with $\hat{V}$. 
Therefore, using Eq.~\eqref{eq:weak-expanded}, 
\bes
\bea
\mathcal{P}_{\epsilon,\hat{V}}\left({\mathcal{U}}(\frac{\tau}{M})\varrho\right) &=& P_{\epsilon,\hat{V}} \left( \sum_{\alpha, \beta \in\{0,1\}} H_{\alpha} \varrho H_\beta \right) \\
&=& (1-\zeta)P_{\infty,\hat{V}} \left(\sum_{\alpha\in\{0,1\}} H_{\alpha} \varrho H_\alpha \right) 
+ \zeta\sum_{\alpha, \beta \in\{0,1\}} H_{\alpha} \varrho H_\beta .
\eea
\ees
The projective measurement $P_{\infty,\hat{V}}$ projects $\varrho'\equiv\left(\sum_{\alpha\in\{0,1\}} H_{\alpha} \varrho H_\alpha \right)$ into $\varrho'_+\equiv P_{+\hat{V}}\varrho' P_{+\hat{V}}/p_+ = \varrho'$ or $\varrho'_-\equiv P_{-\hat{V}}\varrho' P_{-\hat{V}}/p_-= \varrho'$, with probabilities $p_\pm = \textrm{Tr}[P_{\pm\hat{V}}\varrho']=1/2$, with corresponding measurement outcomes $+1$ and $-1$, respectively. Since the measurement is non-selective, the post-(projective-)measurement state is $p_+\varrho'_++p_-\varrho'_- = \varrho'$, so that
\bea
\mathcal{P}_{\epsilon,\hat{V}}\left({\mathcal{U}}(\frac{\tau}{M})\varrho\right) = (1-\zeta)\sum_{\alpha\in\{0,1\}} H_{\alpha} \varrho H_\alpha 
+ \zeta\sum_{\alpha, \beta \in\{0,1\}} H_{\alpha} \varrho H_\beta .
\label{eq:weak-outcome}
\eea
}

\subsection{The simulation}
Next, let us show how the outcome of the many-body weak measurement, Eq.~\eqref{eq:weak-outcome}, can be simulated using single qubit measurements and $2$-qubit gates. First, we introduce an ancilla in a $k$-qubit cat-state $\ket{\Psi_{\textrm{\textrm{cat}},+}}$, where $\ket{\Psi_{\textrm{\textrm{cat}},\pm}} = \frac{1}{\sqrt{2}} \left(\ket{0...0} \pm \ket{1...1} \right)$, and where $k$ is the locality of $\hat{V}$. Let us also introduce controlled-$V_i$ operations, $CV_i$, controlled by the $i$th qubit in the ancilla and targeting the $i$th qubit in the state $\varrho$ we are trying to protect, the ``data state''. The total initial state is $\ket{\Psi_{{\textrm{cat}},+}}\bra{\Psi_{\textrm{\textrm{cat}},+}}\otimes  \varrho$.

By assumption, the noise and control operations on the data do not act on the ancilla, i.e., we replace $\mathcal{U}$ by $\mathcal{I}_{\textrm{cat}}\otimes\mathcal{U}$. Therefore, after Taylor expansion as above, $\ket{\Psi_{{\textrm{cat}},+}}\bra{\Psi_{\textrm{\textrm{cat}},+}}\otimes  \varrho\stackrel{\mathcal{I}_{\textrm{cat}}\otimes\mathcal{U}}{\mapsto}\sum_{\alpha, \beta \in\{0,1\}} \ket{\Psi_{\textrm{\textrm{cat}},+}}\bra{\Psi_{\textrm{\textrm{cat}},+}}\otimes H_{\alpha} \varrho H_\beta$. Rather than applying the many-body measurement $\mathcal{P}_{\epsilon,\hat{V}}$ directly to this state, let us first apply the sequence of $2$-qubit gates $\prod_i CV_i$, which transforms the state into $\sum_{\alpha, \beta \in\{0,1\}} \tilde{H}_{\alpha} \ket{\Psi_{\textrm{\textrm{cat}},+}}\bra{\Psi_{\textrm{\textrm{cat}},+}} \tilde{H}_{\beta} \otimes H_{\alpha} \varrho H_\beta $, where $\tilde{H}_\alpha$ denotes the operation induced on the cat state qubits under the action of the controlled-$V_i$ operators [see Eq.~\eqref{eq:induced}]. Using $CZ_i$, $CX_i = W_i CZ_i W_i${, and $CY_i = C(XZ)_i$} as the controlled-$V_i$ operators, where $W_i$ is the Hadamard gate acting on the $i$th target qubit, operators are transformed as follows:
\bea
\begin{array}{ccccc} 
 CZ_i &:& 
\openone \otimes X &\rightarrow & Z \otimes X\\
  &&  \openone \otimes Z &\rightarrow & \openone \otimes Z\\
 CX_i &:&  
\openone \otimes X &\rightarrow & \openone \otimes X\\
 &&   \openone \otimes Z &\rightarrow & Z \otimes Z\\
{  C{Y}_i }&:& 
\openone \otimes X &\rightarrow & Z \otimes X \\
&&  \openone \otimes Z &\rightarrow & Z \otimes Z\\
 W&:& X &\rightarrow & Z\\
  && Z &\rightarrow & X
\end{array} 
\label{eq:induced}
\eea
{We see that an $X$ or $Z$ error acting on the data (second register) via $H_\alpha$ or $H_\beta$ is always transformed into a $Z$ on the cat (first register), either by $CZ_i$ or $CX_i $. 
Thus, if $\{H_\alpha, \hat V\} =0$ then $\tilde{H}_{\alpha} = \otimes^{{\textrm{(odd)}}}_i Z_i$ (where the product is over the \textit{odd} number of qubits for which $\{H_\alpha, \hat V\} =0$) and hence $\tilde{H}_{\alpha} \ket{\Psi_{\textrm{\textrm{cat}},+}} = \ket{\Psi_{\textrm{\textrm{cat}},-}}$, while if $[H_\alpha, \hat V]=0$ then } { $\tilde{H}_{\alpha} = \otimes^{\textrm{(even)}}_i Z_i$ (where the product is over the \textit{even} number of qubits for which $\{H_\alpha, \hat V\} =0$) and hence $\tilde{H}_{\alpha} \ket{\Psi_{\textrm{\textrm{cat}},+}} =\ket{\Psi_{\textrm{\textrm{cat}},+}}$}.

{At this point we introduce an extra ancilla initialized in $\ket{0}$, execute  a $W$ gate on every qubit of the cat state, then a $\prod_i CX_{i,\textrm{ancilla}}$ gate, where $CX_{i,\textrm{ancilla}}$ is controlled by the $i$th qubit of the cat state and targets the extra ancilla. 
Finally, we apply the weak measurement $\mathcal{P}_{\epsilon, Z}$ just to the extra ancilla, to complete the process.}

{To see how this works, note that the final joint ``simulated state'' right before the measurement of the extra ancilla is
\begin{equation}
\varrho_{\textrm{sim}} \equiv \sum_{\alpha, \beta \in\{0,1\}} X^{s_{\alpha, \hat{V}}} \ket{0}\bra{0} X^{s_{\beta, \hat{V}}} \otimes \tilde{H}_{\alpha} \ket{\Psi_{\textrm{\textrm{cat}},+}}\bra{\Psi_{\textrm{\textrm{cat}},+}} \tilde{H}_{\beta} \otimes  H_{\alpha} \varrho H_\beta,
\end{equation}
where $s_{\alpha,\hat{V}}=1$ if $\{H_\alpha, \hat V\} =0$ or  $s_{\alpha,\hat{V}}=0$ if $[H_\alpha, \hat V] =0$. Using Eq.~\eqref{eq:weak-expanded}, the weak measurement of the ancilla leads to 
\bes
\begin{eqnarray}
\mathcal{P}^{\textrm{ancilla}}_{\epsilon, Z}(\varrho_{\textrm{sim}}) &=& \sum_{\alpha, \beta \in\{0,1\}} \mathcal{P}_{\epsilon, Z} \left( X^{s(\alpha, \hat{V})} \ket{0}\bra{0} X^{s(\beta, \hat{V})}\right) \otimes \tilde{H}_{\alpha} \ket{\Psi_{\textrm{\textrm{cat}},+}}\bra{\Psi_{\textrm{\textrm{cat}},+}} \tilde{H}_{\beta} \otimes  H_{\alpha} \varrho H_\beta\\
&=& (1-\zeta) \sum_{\alpha, \beta \in\{0,1\}} \mathcal{P}_{\infty, Z}  \left( X^{s_{\alpha, \hat{V}}} \ket{0}\bra{0} X^{s_{\beta, \hat{V}}}\right) \otimes \tilde{H}_{\alpha} \ket{\Psi_{\textrm{\textrm{cat}},+}}\bra{\Psi_{\textrm{\textrm{cat}},+}} \tilde{H}_{\beta} \otimes H_{\alpha} \varrho H_\beta \notag\\
 &&+ \zeta  \sum_{\alpha, \beta \in\{0,1\}} X^{s_{\alpha, \hat{V}}} \ket{0}\bra{0} X^{s_{\beta, \hat{V}}} \otimes \tilde{H}_{\alpha} \ket{\Psi_{\textrm{\textrm{cat}},+}}\bra{\Psi_{\textrm{\textrm{cat}},+}} \tilde{H}_{\beta} \otimes  H_{\alpha} \varrho H_\beta \\
 &\equiv& \varrho_{\textrm{sim}}'.
\end{eqnarray}
\ees
Now note that, similarly to the many-body weak measurement case above, 
\bea
\mathcal{P}_{\infty, Z}  \left( X^{s_{\alpha, \hat{V}}} \ket{0}\bra{0} X^{s_{{{\beta}}, \hat{V}}} \right)= \delta_{\alpha,\beta}  X^{s_{\alpha, \hat{V}}} \ket{0}\bra{0} X^{s_{\beta, \hat{V}}}
\eea
Thus, tracing out the extra ancilla and the cat states in the output simulated measurement state leads to the same final state as that of the many-body measurement:
\begin{equation} 
\textrm{Tr}_{\textrm{ancilla,cat}} [\varrho_{\textrm{sim}}'] = (1-\zeta)\sum_{\alpha\in\{0,1\}} H_{\alpha} \varrho H_\alpha + \zeta\sum_{\alpha,\beta\in\{0,1\}} H_{\alpha} \varrho H_\beta
\end{equation}
which is identical to Eq.~\eqref{eq:weak-outcome}, as claimed. 
}

\end{document}